\documentclass[twocolumn,preprintnumbers,amsmath,amssymb, superscriptaddress, showpacs]{revtex4}
\usepackage{graphicx}
\usepackage{hyperref}
\usepackage{fancyheadings}
\begin{document}

\title{\href{http://dx.doi.org/10.1038/ncomms2235}{Extreme sensitivity of graphene photoconductivity to environmental gases}}

\author{Callum J. Docherty}
\affiliation{Clarendon Laboratory, Department of Physics, University of Oxford, Oxford, OX1 3PU, UK}

\author{Cheng-Te Lin}
\affiliation{Institute of Atomic and Molecular Sciences, Academia Sinica, Taipei, 10617, Taiwan}

\author{Hannah J. Joyce}
\affiliation{Clarendon Laboratory, Department of Physics, University of Oxford, Oxford, OX1 3PU, UK}
\author{Robin J. Nicholas}
\affiliation{Clarendon Laboratory, Department of Physics, University of Oxford, Oxford, OX1 3PU, UK}
\author{Laura M. Herz}
\affiliation{Clarendon Laboratory, Department of Physics, University of Oxford, Oxford, OX1 3PU, UK}

\author{Lain-Jong Li}
\affiliation{Institute of Atomic and Molecular Sciences, Academia Sinica, Taipei, 10617, Taiwan}

\author{Michael B. Johnston}
\email{m.johnston@physics.ox.ac.uk}
\affiliation{Clarendon Laboratory, Department of Physics, University of Oxford, Oxford, OX1 3PU, UK}

\pacs{\textbf{Open source version of this article is available at Nature Communications \href{http://dx.doi.org/10.1038/ncomms2235}{DOI: 10.1038/ncomms2235} }}

\begin{abstract}
Graphene is a single layer of covalently bonded carbon atoms that was discovered only eight years ago and yet has already attracted intense research and commercial interest. Initial research focused on its remarkable electronic properties, such as the observation of massless Dirac Fermions and the half-integer quantum Hall effect. Now, graphene is finding application in touch-screen displays, as  channels in high-frequency transistors, and in graphene-based integrated circuits.  The potential for using the unique properties of graphene in terahertz-frequency electronics is particularly exciting, however initial experiments probing the terahertz frequency response of graphene are only just emerging. Here we show that the photoconductivity 
of graphene at terahertz frequencies is dramatically altered by the adsorption of atmospheric gases, such as nitrogen and oxygen. Furthermore, we observe the signature of terahertz stimulated emission from gas-adsorbed graphene.  Our findings highlight the importance of environmental conditions on the design and fabrication of high-speed graphene-based devices.
\end{abstract}

\maketitle

\section*{}
The experimental realisation of graphene, a single layer of graphite, in 2004 \cite{Novoselov04-1892} rapidly led to the observation of massless Dirac fermions and the half-integer quantum Hall effect\cite{Novoselov05-1875}. Since then many potential applications have been suggested\cite{Geim09-2961,Bao12},  and  recently graphene-based touch-screen displays\cite{Bae10-2571}, transistors\cite{Lin09-2977, Wu11-2978}, and integrated circuits\cite{Lin11-2979} have been demonstrated. The high mobility of charge carriers in graphene makes it an ideal material for extremely high-speed devices\cite{Lin09-2977, Wu11-2978}, indeed terahertz frequency switching has been predicted\cite{Geim09-2961}.  The potential for the realisation of such devices was greatly enhanced by the emergence of large area graphene sheets grown by chemical vapour deposition (CVD) \cite{Li09-3420}.  However, the design and production of these high-frequency devices requires a fuller understanding of both nonequilibrium carrier dynamics in graphene \cite{Wang10-2495}, and its susceptibility to environmental conditions \cite{Klarskov11-3367}.

Ultrafast spectroscopic techniques provide an excellent probe of non-equilibrium carrier dynamics in graphene\cite{Docherty12-3608}. Ultrafast cooling and relaxation of photoexcited charge carriers in graphene have been observed using optical pump--optical probe spectroscopy. Several studies \cite{Dawlaty08-1882,Hale11-3267,Wang10-2495} have shown that graphene becomes more transparent at optical frequencies after intense photoexcitation (photoinduced bleaching), owing to state blocking of interband absorption. Furthermore these studies showed that thermalisation of photoexcited carriers occurs within tens of femtoseconds, and a subsequent cooling by carrier--phonon interactions occurs on a picosecond timescale. 

While optical photons probe interband charge dynamics, terahertz frequency photons are excellent probes of intraband dynamics.  Studies of intraband relaxation in graphene have been performed using optical pump--terahertz probe spectroscopy (OPTPS) \cite{Docherty12-3608, George08-1714, Choi09-1876, Strait11-3193}.  George et al.\cite{George08-1714} revealed that the absorption of the terahertz photons in epitaxially grown graphene increases after intense (interband) optical excitation.  The increase in absorption was attributed to an increase in the density of mobile charge carriers after photoexcitation. However, a comparable study on chemical vapour deposition (CVD) grown graphene found the opposite effect, photoinduced bleaching \cite{Hwang11}. It is known that CVD grown graphene is p-doped, due to residual impurities and defects from the production process \cite{Horng11-2856}, unlike epitaxial graphene \cite{Choi09-1876}. Despite this, equilibrium carrier dynamics in CVD grown graphene have been shown to be similar to other forms of graphene in the terahertz regime, suggesting that growth conditions cannot explain the discrepancy between the epitaxial and CVD graphene in pump--probe studies \cite{Lee11-3261}.

In this study, we utilised terahertz time domain spectroscopy to measure the photoconductivty of CVD grown graphene in four different environment types: air, oxygen, nitrogen and vacuum. We demonstrate that CVD grown graphene can display either photoinduced bleaching or photoinduced absorption depending on the type and density of gas adsorbed on its surface. Our results highlight the dramatic effects environmental conditions have on the high-frequency electrical properties of graphene.  These results have important implications for the design of future graphene electronic devices. In particular, the order-of-magnitude changes in photoconductivity we observe could provide a basis for precise graphene based gas monitors. Conversely, the sensitivity of the optoelectronic properties of graphene to environmental conditions will need to be addressed during the design and fabrication of future devices.

\section*{Results}

\subsection*{Sample preparation and characterisation}
\begin{figure}
\centering
\includegraphics[width=8.6cm]{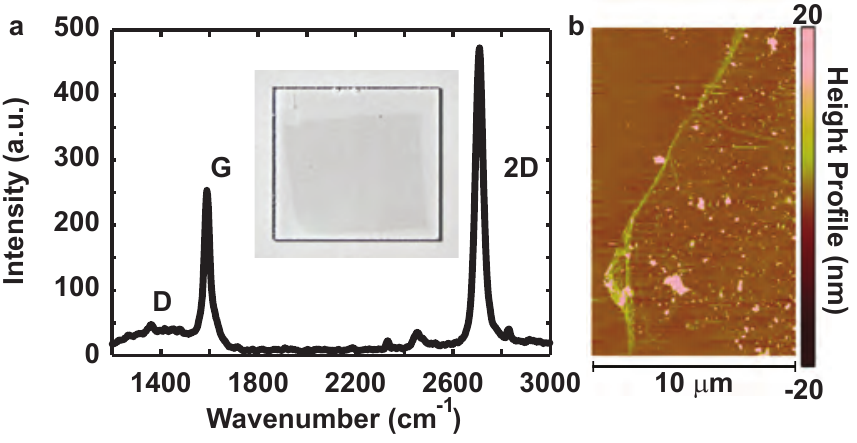}
\caption{\label{Raman}Characterisation of graphene samples. \textbf{a} Raman spectroscopy. The relative intensities of the 2D and G peaks suggest monolayer graphene \cite{Ferrari06-1995}. \textbf{Inset} Photograph of graphene sheet with area of $1\times1$\,cm$^{2}$ on quartz. \textbf{b} Surface topography of CVD-grown graphene measured by AFM. The residuals remaining on graphene surface can be attributed to incomplete decomposition or carbonization of PMMA.}
\end{figure}

The CVD samples used in this study were grown on copper foil, and transferred to  z-cut quartz substrates using poly(methyl methacrylate)  as a supporting layer (for details see the Methods section). Raman microscopy and atomic force microscopy (AFM) revealed that the samples were primarily a graphene monolayer with relatively few defects. A typical Raman spectrum is shown in Figure \ref{Raman} and exhibits a Lorentzian 2D Raman peak (2700\,cm$^{-1}$), with the ratio of intensities of the 2D and G (1580\,cm$^{-1}$) peaks, I$_{\mathrm{2D}}$/I$_{\mathrm{G}} \approx 2$, which is indicative of monolayer growth. Additionally, the relative weakness of the D peak (1350\,cm$^{-1}$) demonstrates the small number of defects and low disorder in the samples, confirming their high quality \cite{Ferrari06-1995}. UV-visible absorption measurements revealed an absorption of $\sim\pi\alpha$, where $\alpha$ is the fine structure constant, also suggesting monolayer growth (see Supplementary Methods) \cite{Nair08-1994}. AFM images (Figure \ref{Raman}b) showed the graphene sheet thickness measured from the edge to be $\sim 0.72$\,nm, further suggesting monolayer growth \cite{Gupta06-3369}. Hall measurements revealed the sample to be p-doped, with hole density $8\times10^{12}$\,cm$^{-2}$, and Fermi level $\varepsilon_{\rm F}\sim-0.3$\,eV.

\subsection*{Terahertz time domain spectroscopy}
In order to assess the influence of atmospheric gas adsorption on graphene we utilised terahertz spectroscopy. This technique allowed us to observe intraband charge dynamics and investigate the high-frequency electrical properties of graphene. Terahertz spectroscopy is ideal for electrically characterising surface sensitive materials such as graphene as it is a non-contact technique, and covers an important  frequency range ($\sim$ 100\,GHz -- $>$3THz) for high speed device operation\cite{Ulbricht11-2937}. OPTPS measurements were performed with CVD-grown graphene samples in an atmosphere of air, N$_{{2}}$ or O$_{{2}}$, or in a vacuum. Before each set of measurements, the sample was left under a vacuum of 10$^{-5}$\,mbar for two hours, to remove existing adsorbants (see Supplementary Methods for Langmuir coverage calculations). For the vacuum measurements, the graphene was left for a further two hours with 800\,nm (1.55\,eV) ``pump'' photons incident upon the sample before measurement. For the gaseous measurements, 1\,bar of air, N$_{{2}}$ or O$_{{2}}$ was introduced after two hours under vacuum, and measurements were taken after a further two hours, again with the ``pump'' beam incident upon the sample. Measurements were found to be reproducible on different samples and were independent of sample history. That is, the effect of exposure to any of the gases could be ``reset'' by placing the sample in vacuum for 2 hours.  

\begin{figure*}
\centering
\includegraphics[width=17cm]{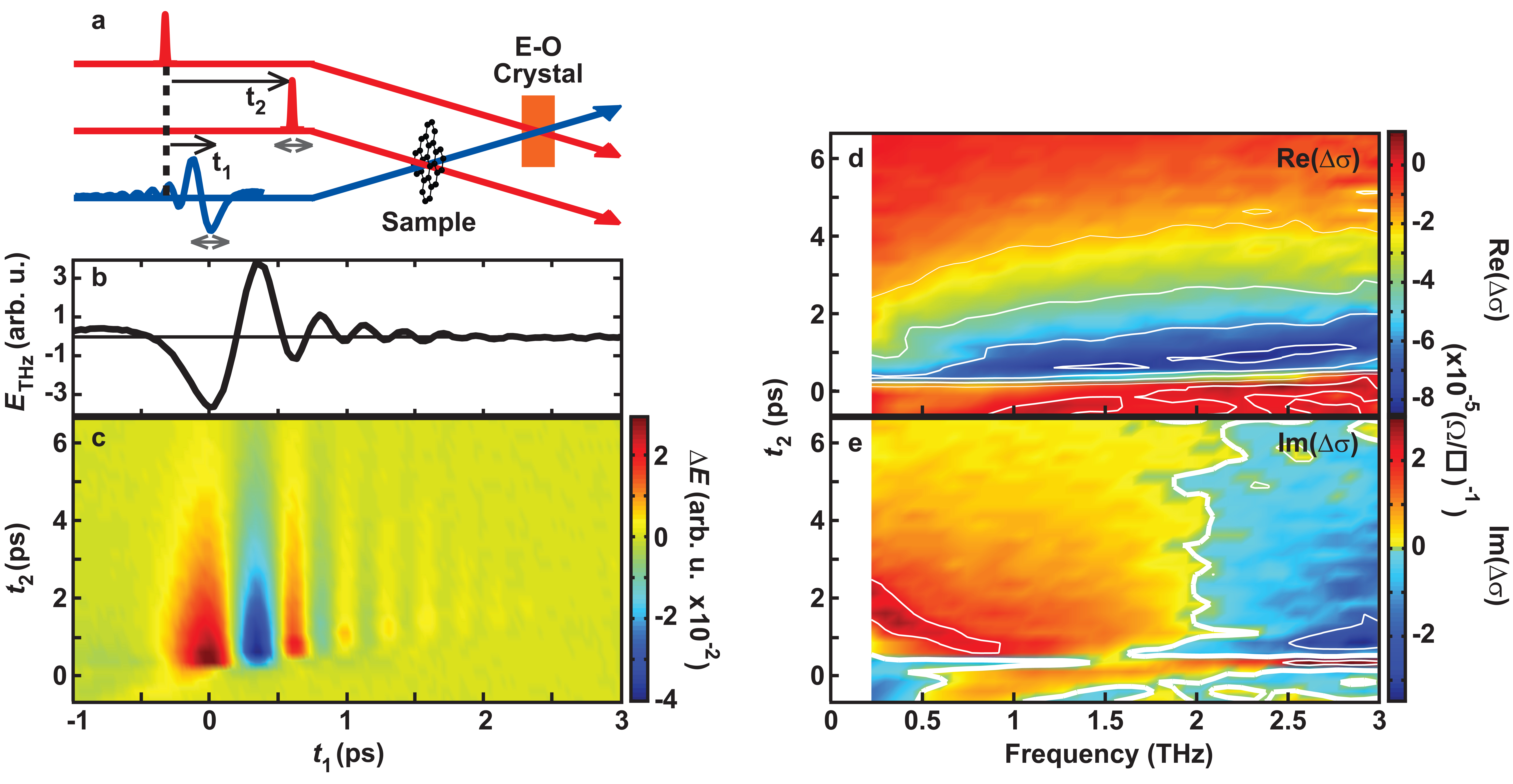}
\caption{\label{O2}Terahertz spectroscopy of graphene in O$_{2}$. \textbf{a} Schematic representation of the measurement. Varying $t_{1}$ allows the electric field, $E_{\mathrm{THz}}$, of the terahertz probe pulse to be sampled in the electro-optic (E-O) crystal. The sample can be probed at different times after photoexcitation by altering t$_{2}$. \textbf{b,c} terahertz electric field, $E_{\mathrm{THz}}$, transmitted through the sample (\textbf{b}), and the corresponding photoinduced change in transmission, $\Delta E$ (\textbf{c}), as a function of pump--probe delay time, $t_{2}$. \textbf{d,e}  Real (\textbf{d}) and imaginary (\textbf{e}) parts of the frequency dependent photoconductivity as a function of $t_{2}$. The contours highlight increments of $2\times10^{-5}$\,S$\square$. The thicker contour in \textbf{e} represents Im($\Delta\sigma) = 0$, corresponding to the Lorentzian peak frequency, $\omega_{0}$.}
\end{figure*}

For the measurements, the electric field of a terahertz pulse transmitted through the graphene was sampled after photoexcitation by an 800\,nm, 180\,$\mu$J/cm$^{2}$ fluence pump pulse, $E^{\mathrm{ON}}_{\mathrm{THz}}$, and without the photoexcitation, $E^{\mathrm{OFF}}_{\mathrm{THz}}$. The photoinduced change in terahertz transmission of the sample, $\Delta E=E^{\mathrm{ON}}_{\mathrm{THz}}-E^{\mathrm{OFF}}_{\mathrm{THz}}$ was recorded using lock-in detection. The principle of the measurements is illustrated in Figure \ref{O2}a. The terahertz pulse probes the photoexcited graphene sample at a time $t_{2}$ after photoexcitation. The terahertz probe is then mapped out in time by varying the difference in time of arrival, $t_{1}$, between the terahertz pulse and an 800\,nm gate pulse at an electro-optic detection crystal (GaP).

In Figures \ref{O2}b and c we present the measured terahertz electric field, $E_{\mathrm{THz}}$, transmitted through graphene in an pure oxygen atmosphere, and the corresponding change in transmission due to photoexcitation, $\Delta E$. The induced $\Delta E$ decreases as the delay between pump and probe, $t_{2}$, is increased.  The photoinduced change in terahertz transmission is seen to decay within $\sim$5\,ps.  Further information may be extracted from these data by performing a Fourier transform on the electric field data to give directly the frequency dependent complex photoconductivity, $\Delta\sigma(\omega)$, without resorting to the Kramers-Kronig relations, via \cite{Choi09-1876}

\begin{eqnarray}
\centering
\Delta\sigma(\omega)= -\left(\frac{1+n_{\mathrm{s}}}{Z_{0}}\right)\frac{\Delta E(\omega)}{E_{\mathrm{THz}}(\omega)},
\label{FTeq}
\end{eqnarray}

where  $Z_{0}$ is the impedance of free space, $Z_{0}=377\,\Omega$, and $n_{\mathrm{s}}$ is the refractive index of the quartz substrate, $n_{\mathrm{r}}=2.1$ \cite{Grischkowsky90-261} at terahertz frequencies.
Figures \ref{O2}d and e show the real and imaginary parts of the photoconductivity respectively as a function of pump--probe delay time.

\begin{figure}
\centering
\includegraphics[width=8.6cm]{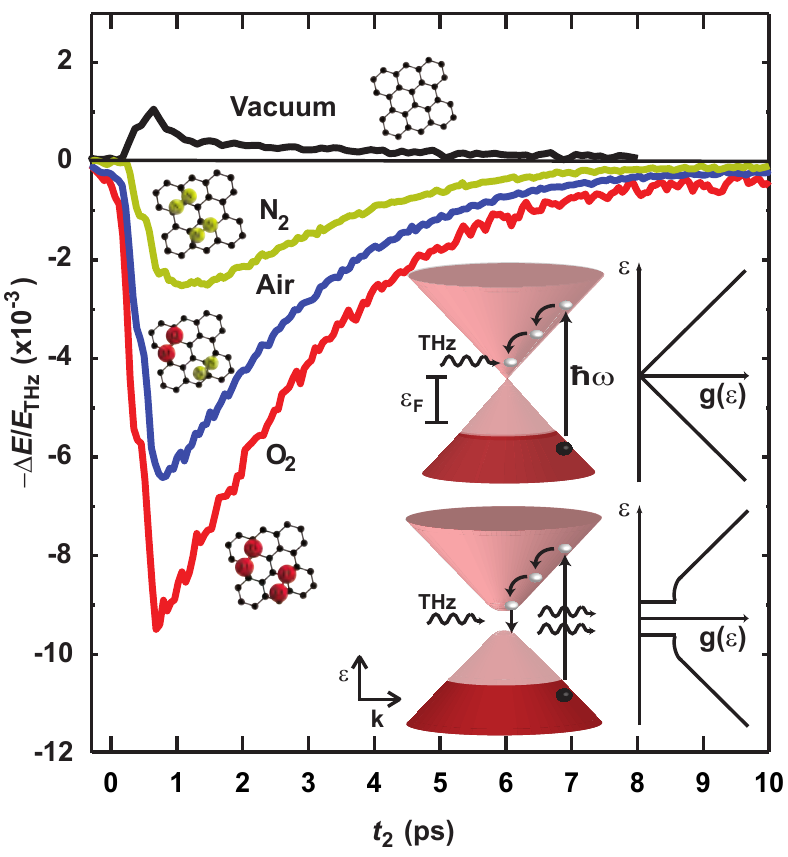}
\caption{\label{pumpcomp} Environmental dependence of pump induced change in terahertz photoconductivity, -$\Delta E /E_{\mathrm{THz}}$. Here, $\Delta E=E^{\mathrm{ON}}_{\mathrm{THz}}-E^{\mathrm{OFF}}_{\mathrm{THz}}$. From top to bottom: graphene in vacuum, N$_{{2}}$, air and O$_{{2}}$. Photoinduced absorption is observed in vacuum, but photoinduced bleaching is observed in gaseous environments. Inset: Schematic representation of the pump--probe measurements of p-doped graphene ($\varepsilon_{\rm F}\sim-0.3$\,eV) in vacuum (top), and in the presence of gases (bottom), with corresponding density of states ($g(\varepsilon$)). 
The high energy pump ($\hbar\omega$) excites carriers far above the Dirac point, which cool by phonon emission. In gapless graphene, terahertz photons can be absorbed by the photoexcited carriers. In the presence of gases, stimulated emission releases extra terahertz photons, yielding a bleaching pump-probe signal.}
\end{figure}

To more easily compare the photoconductivity in different atmospheres, we take slices of the data from Figure \ref{O2}, and the corresponding data for the other environment types. For instance, holding $t_{1}$ constant at the maximum terahertz electric field (e.g. $t_{1}$=0 in Figure \ref{O2}c) and observing the resulting $\Delta E$ as pump--probe delay time $t_{2}$ is varied, yields an optical pump--terahertz probe measurement, which provides the average terahertz response of the sample. In Figure \ref{pumpcomp}, we plot -$\Delta E / E_{\mathrm{THz}}$, which is proportional to the photoconductivity of the graphene, as a function of the delay time between the pump and probe pulses in the four atmosphere types. It should be noted that due to the large number of holes already present near the Dirac point in this p doped sample, the transient response probed by the terahertz is dominated by electron dynamics.

Immediately after photoexcitation in vacuum, the terahertz conductivity increases (photoinduced absorption) within 1\,ps, and subsequently relaxes with two distinct time scales. Initially fast cooling occurs within $\sim$0.6\,ps, and a longer, exponential cooling occurs with a time scale of $\sim$2.5\,ps. Such a relaxation is consistent with other studies, which associated the first rapid cooling with emission of optical phonons, followed by a bottleneck when the carrier temperature cools below the hot optical phonon temperature ($\sim0.196$ and 0.162\,eV \cite{Strait11-3193}) , slowing further cooling \cite{Strait11-3193, Wang10-2495, Hale11-3267}.

However, when graphene is placed in a gaseous atmosphere, the photoinduced change is dramatically different. Surprisingly, the photoconductivity signal flips from being purely positive in vacuum to purely negative in all the gas environments. Following photoexcitation in air, N$_{\mathrm{2}}$ or O$_{\mathrm{2}}$, the terahertz conductivity decreases within 1\,ps, and then relaxes with a common monoexponential time scale of 2.5\,ps. The similarity between this time constant and that of cooling in vacuum suggests the continued importance of carrier--phonon interactions.  Adsorbed O$_2$ was found to have the strongest influence on charge dynamics is graphene, with the negative photoconductivity almost an order of magnitude greater than the induced absorption under vacuum.  A negative transient absorption, such as we observe, may be attributed to either stimulated emission in the samples or a bleaching effect. To understand more about the origin of the negative photoconductivity it is informative to consider the photoconductivity spectra at a fixed time after photoexcitation.  

\begin{figure}
\centering
\includegraphics[width=8.6cm]{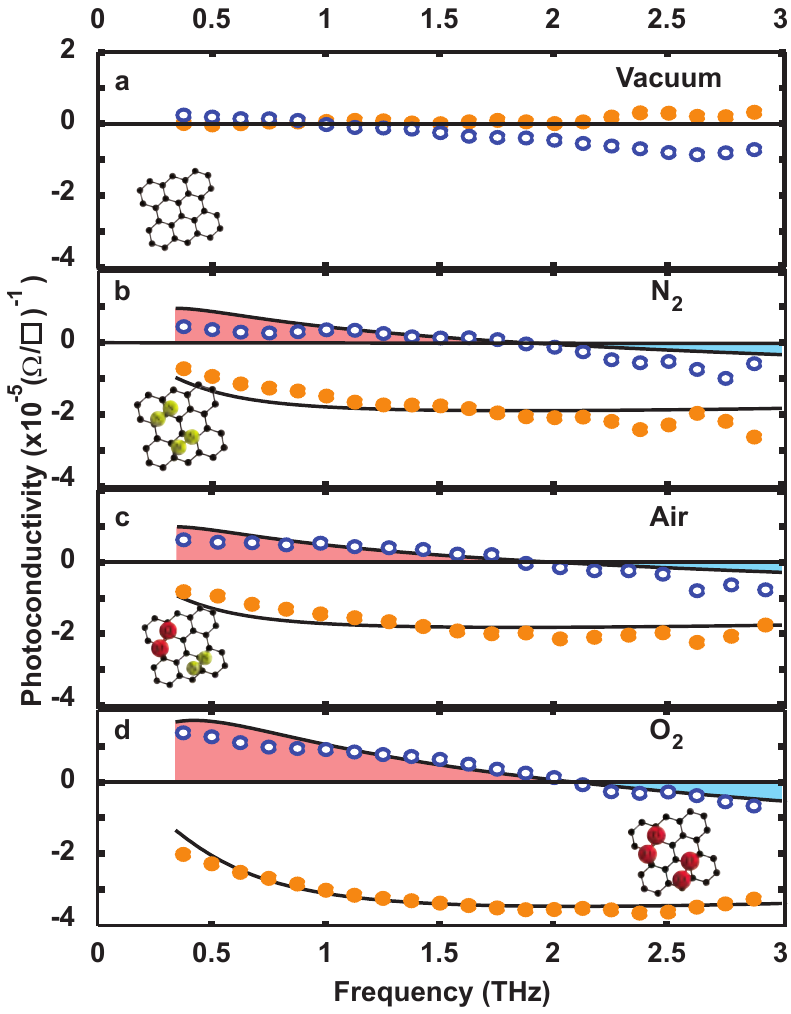}
\caption{\label{spectral}Photoconductivity spectra of graphene in different environments. \textbf{a-d} Photoconductivty in vacuum (\textbf{a}), N$_{{2}}$ (\textbf{b}), air (\textbf{c}) and O$_{{2}}$ (\textbf{d}), taken at 2\,ps after photoexcitation by the optical pump. Solid orange dots show the real part of the photoconductivity, hollow blue dots show the imaginary part. The solid black line in \textbf{b-d} is a Lorentzian model fit to the conductivity data, demonstrating the Lorentzian form the of photoconductivity in gaseous environments. Shading is used to emphasise the change from positive to negative Im($\Delta\sigma$). This point corresponds to the peak Lorentzian frequency, $\omega_{0}$.}
\end{figure}

Figure \ref{spectral} shows the photoconductivity in all four atmosphere types at 2\,ps after photoexcitation ($t_{2}=2$\,ps).  The spectrum of graphene in vacuum reveals a relatively flat and positive real part of photoconductivity, in stark contrast to the strongly negative (real) photoconductivity seen for gas-absorbed graphene.  Furthermore the spectral photoconductivity of graphene samples in the presence of N$_2$, air and O$_2$ can be seen to exhibit a Lorentzian form,
\begin{eqnarray}
\centering
\sigma_{\mathrm{Lorentz}} \propto \frac{i \omega}{\omega^{2} - \omega_{0}^{2} + i\omega\Gamma},
\label{Lorentz}
\end{eqnarray}

where $\Gamma$ is the linewidth, and $\omega_{0}$ is the resonant frequency. The solid lines in Figure \ref{spectral} represent a least square fit of Eq. \ref{Lorentz} to the experimental data. The fitted resonant frequency for graphene in air and N$_{{2}}$ was found to be $\omega_0 \sim 1.8$\,THz, compared with $\omega_0 \sim 2$\,THz for graphene in an O$_{{2}}$ environment. The resonance in all gas types is spectrally very broad; the fitted linewidth was $\Gamma \approx 10$\,THz, independent of the gas type. The Lorentzian fit begins to deviate from the measured photoconductivity at low frequencies.  The deviation is most noticeable for O$_2$-exposed graphene (Figure \ref{spectral}d). The non-zero value of Re($\Delta\sigma$) evident in the data at low frequency suggests an underlying Drude-like response, however inclusion of a Drude term does not enhance the fit within our observable bandwidth, so has been neglected.  In contrast, the photoinduced absorption in vacuum leads to a small, positive, spectrally flat real conductivity, with a non-negligible imaginary part. 

A negative transient absorption signal with a Lorentzian lineshape is characteristic of stimulated emission.  Hence, our data provide evidence that stimulated emission of terahertz photons is occurring in photoexcited graphene in the presence of atmospheric gases.  Theoretical studies have predicted amplified stimulated emission of terahertz radiation from sufficiently intensely photoexcited graphene\cite{Ryzhii07-1037}. The proposed mechanism is that after photoexcitation, the charge carriers relax and populate states around the Dirac point, from where they can radiatively recombine. Thus terahertz emission over a large spectral range is expected to be possible\cite{Ryzhii07-1037}. Whilst our measurements do not show amplification of the terahertz probe (losses are still greater than gain), stimulated terahertz emission, below the gain threshold is consistent with our observations.

\section*{Discussion}

We now consider the mechanism for the dramatic changes of photoconductivity in graphene under the different gas environments.  The ability to ``reset'' the effects of gas exposure by placing the graphene samples in vacuum for two hours (at room temperature) suggests that the gas molecules are physisorbed rather than chemisorbed to the graphene surface. Physisorption of O$_2$ and N$_2$ on graphene could act as a dopant \cite{Shi09-2865}, lead to the opening of a small bandgap at the Dirac point \cite{Ito08-3682, Berashevich09-3421} , and/or lead to a modification of the density of states \cite{Dai10-3401, Ando09-3678}.  Hall measurements (see Methods) showed only ~3\% changes in the Fermi level after gas adsorption, indicating that doping alone is unlikely to account for the huge change in photoconductivity we observe. The opening of a $\sim 8$\,meV bandgap upon gas adsorption, and associated increase in the density of electron and hole states near the band extrema  would be consistent with our observations.  In this case, electrons injected into the conduction band by 800\,nm (1.55\,eV) photoexcitation would relax rapidly to near the band minimum via phonon mediated processes, as shown on the inset of Figure \ref{pumpcomp}. Recombination for the photoinjected electrons across the bandgap with a sea of holes would lead to the observed stimulated emission signal.  The spectral broadness  ($\Gamma \approx 10$\,THz) of the observed transition may then be explained by a combination of the following factors: a high interband transition rate (lifetime broadening), fast dephasing via electron-electron interactions and a cooling bottleneck below the optical phonon energies.  However full \emph{ab-inito} calculations would be required to confirm the exact mechanism.  Note that since CVD graphene is heavily p-doped, the opening of a bandgap is not directly observable by terahertz transmission spectroscopy, as there are no electrons near the band gap to be excited by the low energy terahertz photons (see Supplementary Discussion and inset of Figure \ref{pumpcomp}).

The adsorption of molecules on graphene would locally open a bandgap and hence alter the otherwise linear density of states ($g(\varepsilon)$) of the graphene, thereby increasing $g(\varepsilon)$ at the top of the valance band and bottom of the conduction band (see inset of Figure \ref{pumpcomp}) \cite{Ando09-3678}.  However, a gap would only be opened in regions around adsorbed gas molecules, creating islands of semiconducting graphene amongst the unaltered, gapless graphene (see Supplementary Methods for Langmuir coverage calculations). As the terahertz probe is much larger than these puddles, the observed photoconductivity behaviour is the result of an interaction with of both types of graphene. Increased stimulated emission will therefore be observed if gas molecules are more effectively adsorbed on the surface, thus creating a higher proportion of semiconducting material. Of the gases used in this study, O$_2$ is most likely to interact with graphene  \cite{Ryu10-3685, Sato11-2936} , which is consistent with the increased negative photoconductivity signal observed in pure O$_2$ compared with air and N$_2$.

There are a number of other possible explanations for a Lorentzian photoconductivity in graphene. A positive Lorentzian conductivity may originate from the excitation of plasmons \cite{Parkinson09-1682}. Terahertz surface plasmons have been predicted theoretically \cite{Rana08-1889, Ryzhii07-1900}, and recently observed in spatially confined graphene \cite{Ju11-3284}. However, the peak frequency of a plasmon dependent conductivity varies with carrier concentration as $\omega_{0} \propto n^{1/4}$ \cite{Ju11-3284} and Re($\Delta\sigma$) is expected to be positive, in contrast to our experiments. Figure \ref{O2} demonstrates that our measured $\omega_{0}$ does not decrease with time after photoexcitation, and hence is independent of photoexcited carrier concentration. Additional measurements showed $\omega_{0}$ to be independent of pump fluence (see Supplementary Discussion). Thus it is unlikely that excitation of plasmons explains the Lorentzian response.  A negative photoconductivity may also be attributed to a photo-induced reduction in conductivity, for example by an increase in the carrier--carrier or carrier--phonon (intraband) scattering rates.  However it is not clear why such mechanisms would produce a Lorentzian form, or occur only in gas adsorbed graphene.  Finally, ground-state bleaching could lead to a negative photoconductivity with Lorentzian form,  however the states associated with the absorption of (1.55\,eV) optical photons in graphene are well separated from the states near the Dirac point that are associated with interband terahertz absorption and emission (see the inset of Figure \ref{pumpcomp}). We therefore conclude that the opening of a small band gap in the presence of adsorbed gas molecules is the most likely explanation for our measurements.

\section*{}
In summary, we have used terahertz time domain spectroscopy to noninvasively probe photoexcited carrier dynamics in CVD grown graphene. We have observed that gas adsorption can dramatically alter the high frequency electrical response of graphene sheets. We also provide experimental evidence for stimulated emission of terahertz photons in graphene. Furthermore our results demonstrate the huge influence environmental factors
can exert on the behaviour of hot carriers in CVD graphene, and highlight the importance of consideration of such factors for future CVD graphene opto-electronic devices.

\section*{Methods}

\subsection*{Synthesis and Transfer of CVD Graphene}
Large-area graphene films were prepared by chemical vapour deposition on 25\,$\mu$m copper foil (Alfa Aesar). Prior to the growth, the copper foil was heated in hydrogen from room temperature to 1000\,$^{\circ}$C for 60 minutes for removal of surface oxides. As temperature reached 1000\,$^{\circ}$C, a mixed flow of H$_{{2}}$ (15\,sccm) and CH$_{{4}}$ (60\,sccm) was introduced for graphene growth. After CVD, the graphene film was separated from copper foil and then attached to the z-cut quartz substrate by the polymer transfer process: Poly(methyl methacrylate) (MicroChem Co., NANO� PMMA 950K A4) was spin-coated on graphene/copper foil to form a protective layer. The copper foil was etched at 25\,$^{\circ}$C in ferric nitride solution (50\,g/L, J.T.Baker ACS reagent, 98\%), leaving a transparent PMMA/graphene film floating on the solution. After rinsing, the film was transferred onto the quartz substrate, and the PMMA capping layer was dissolved in acetone at 60\,$^{\circ}$C overnight. In order to remove the residual PMMA, graphene films were annealed at 450\,$^{\circ}$C at 500\,Torr in the mixed gases of H$_{{2}}$ (20\,sccm) and Ar (80\,sccm) for decomposition of the polymer.

\subsection*{Characterisation}
Primary characterisation of the sample was performed by Raman spectroscopy \cite{Ferrari06-1995}, using a NT-MDT confocal Raman microscopic system (laser wavelength 473\,nm, power 0.5-1\,mW, spot size 0.5\,$\mu$m). 

Additional characterisation was performed using UV-visible absorption spectroscopy (Perkin-Elmer Lambda 9 UV-visible-
NIR Spectrophotometer) in the range 500-1300\,nm, confirming the expected constant $\pi\alpha$ absorption of a single layer of graphene in this wavelength range \cite{Nair08-1994}. 

Nominally identical samples were characterised using atomic force microscopy (Veeco Dimension-Icon) and room temperature Hall effect measurements based on the Van der pawl method (3S Co., NI PXI-1033). From the Hall effect measurements averaged across 8 samples, the sheet resistance, carrier density and Hall mobility were found to be $1100 \pm 350$\,$\Omega /\square$, $8 \pm 2 \times10^{12}$\,cm$^{-2}$ and $800 \pm 250$\,cm$^{2}$/Vs respectively. The characterizations show that the quality of the graphene samples is fairly good.

Further Hall effect measurements were performed to examine the equilibrium electrical properties of the graphene samples under different gaseous environments.  
Two systems were used for the measurements, one based on a permanent magnet (0.68\,T), and another using a swept electromagnet (0 -- 0.91\,T).  Samples were held in a vacuum vessel, which could be purged with different gasses. The Fermi levels were deduced using the relation $\varepsilon_{\mathrm{F}}=\hbar v_{\mathrm{F}} \sqrt{\pi n}$, where $v_{\mathrm{F}} \approx 10^{6}$\,ms$^{-1}$ \cite{Novoselov05-1875, Ju11-3284}.

The equilibrium carrier density, and hence Fermi level, were found to be relatively insensitive to the gaseous environment, with variations in $n$ and $\varepsilon_{\rm F}$ between samples being greater than changes induced by the gases.  It was found that, compared with vacuum conditions, exposure to pure nitrogen gas decreased the carrier density by $\sim5\%$, exposure to oxygen increased the carrier density by $\sim7\%$ and exposure to air increased it by $\sim3\%$, correspnding to Fermi level changes of $\sim2\%$, $\sim3\%$ and $\sim1\%$ respectively.  These changes are consistent with previous observation of oxygen and nitrogen exposure doping CVD graphene p-type and n-type respectively \cite{Shi09-2865, Daas12}.

\subsection*{Terahertz spectroscopy}

A Ti:sapphire regenerative amplifier (Spectraphysics Spitfire Pro) was used as the pulse laser source for both terahertz generation and optical pumping of the sample: 800\,nm, 45\,fs pulse duration, 5\,kHz repetition rate, 4\,W average power. 2.9\,W of this beam, reduced to 0.29\,W with an neutral density filter, was used to photoexcite the graphene sample, with a maximum fluence of 180\,$\mu$J/cm$^{2}$ incident upon the sample, and beam width 2\,mm. 

A fraction (200\,$\mu$J per pulse) of the amplifier beam was used to generate the terahertz probe beam by optical rectification \cite{Ulbricht11-2937} in a 2\,mm GaP crystal. The terahertz beam had a width of 0.48\,mm at the sample position. The area of graphene probed by the terahertz beam was much smaller than the area photoexcited by the pump beam, thus the probe measured an area of roughly constant photoexcited carrier density.

The remainder of the amplifier beam (16\,$\mu$J per pulse) was used as a gate beam for electro-optic sampling \cite{Ulbricht11-2937} of the terahertz field in a 200\,$\mu$m GaP crystal. The terahertz electric field, $E_{\mathrm{THz}}$, was detected by a balanced photodiode circuit, and the signal extracted using a lock-in amplifier (Stanford Research SR830), referenced to a 2.5\,kHz chopper in the terahertz generation beam. A further lock-in (SR830) was used to detect the optical pump induced change in $E_{\mathrm{THz}}$, $\Delta E$, by referencing to a 125\,Hz chopper in the sample pump beam.

The gaseous atmospheres used in the terahertz measurements were ambient air, 99.998\% pure oxygen free N$_{2}$ and 99.99\% pure O$_{2}$. A vacuum of 10$^{-5}$\,mbar, generated using an oil--free pumping system (turbo molecular pump and scroll pump), was used both for the vacuum measurements, and to remove adsorbants from the sample before gaseous measurements.

\section*{Acknowledgements}

The authors would like to thank the EPSRC (UK) for financial support.

\section*{Author Contributions}
CJD performed the THz measurements and data analysis.  CTL and LJL grew and processed the graphene samples and performed Raman spectroscopy.  CJD, HJJ and RJN performed Hall measurements in Oxford.  CTL and LJL performed Hall measurements in Taipei.  MBJ conceived the study, while all authors discussed the results.  CJD and MBJ wrote the manuscript supported by LMH and RJN.  

\section*{Competing financial interests}
The authors declare that they have no competing financial interests.

\end{document}